\documentstyle[12pt]{article}

\def\Journal#1#2#3#4{{#1} {\bf #2}, #3 (#4)}

\def\PRA{{\em Phys. Rev.} A}

\def\CPAM{\em Comm. Pure Appl. Math.}
\def\PRS{{\em Proc. Roy. Soc. (London)} A}

\def\be{\begin{equation}}
\def\ee{\end{equation}}
\def\bea{\begin{eqnarray}}
\def\eea{\end{eqnarray}}

\title{ALGEBRAIC MEAN FIELD THEORY}

\author{Ts.\ Dankova and  G.\ Rosensteel \\ 
Physics Department, Tulane University, \\
 New Orleans, Louisiana 70118 }

\begin{document}

\maketitle

\abstract{Mean field theory has an unexpected group theoretic mathematical foundation. Instead of representation theory which applies to most group theoretic quantum models, Hartree-Fock and Hartree-Fock-Bogoliubov have been formulated in terms of coadjoint orbits for the groups $U(n)$ and $O(2n)$. The general theory of mean fields is formulated for any arbitrary Lie algebra {\textbf g} of fermion operators. The moment map provides the correspondence between the Hilbert space of microscopic wave functions and the dual space {\textbf g}$^\ast$ of densities. The coadjoint orbits of the group in the dual space are phase spaces on which time-dependent mean field theory is equivalent to a classical Hamiltonian dynamical system. Indeed it forms a finite-dimensional Lax system. The $SU(3)$ mean field theory is constructed explicitly in the coadjoint orbit framework.}

\section{Introduction}
There are two basic types of models widely used in describing an atomic nucleus: shell model and mean field theory. These theories are the foundations of many-body science. Both of them can be built on a Lie algebra of fermion operators with the differences arising from the way the algebra determines the model states. In a shell model the states of the many-fermion system form an irreducible unitary representation of the Lie algebra. The Hamiltonian is an element of the enveloping algebra and its eigenvalues form the energy spectrum. The matrix elements of the Lie algebra generators give the transition rates between the states. In a mean field theory the model states are restricted to an orbit of the corresponding Lie group in Fock space. This orbit is a general surface and it is not a vector space. There is a well-defined mapping from the orbit to the space of densities. Its image is a phase space and time-dependent mean field theory is a classical dynamical system on it.

A familiar illustration of this idea is the algebra $u(n)$ of one body operators $\hat{X} = \sum _{\alpha \alpha^{\prime}} X_{\alpha \alpha^{\prime}} a^{\dag}_{\alpha} a_{\alpha^{\prime}}$, where $a^{\dag}_{\alpha}$, $a_{\alpha^{\prime}}$ are the fermion creation and annihilation operators and $\alpha,\alpha^{\prime} = 1,\ldots,n$ index an orthonormal basis for the $n$-dimensional single particle space. The Hilbert space of an A-fermion system is the totally antisymmetric irreducible unitary representation of $u(n)$ of dimension $n!/A!(n-A)!$. The shell model states are eigenfunctions of the Hamiltonian 
\begin{equation}
H = \sum_{\alpha \alpha^{\prime}} T_{\alpha \alpha^{\prime}} a^{\dag}_{\alpha} a_{\alpha^{\prime}} + \frac{1}{4} \sum_{\alpha \beta \alpha^{\prime} \beta^{\prime}} V_{\alpha \beta \alpha^{\prime} \beta^{\prime}} a^{\dag}_{\alpha} a^{\dag}_{\beta} a_{\beta^{\prime}} a_{\alpha^{\prime}} .
\end{equation}

In the case of Hartree-Fock mean field theory, the many-body states are approximated by Slater determinants that form an orbit of the group $U(n)$ in Fock space.~\cite{RR81}  The group of unitary basis transformations for the single particle wave function space is $U(n)$. A determinant $\psi_{HF}$ is typically a poor approximation to the shell model eigenstate $\psi$. Indeed these two wave functions are almost orthogonal,  $\langle \psi \mid \psi_{HF} \rangle \approx 0$. But the expectation values of the one-body operators are similar whether taken with respect to the shell model ground state or its mean field approximation, 
\begin{equation}
\langle \psi \mid \hat{X} \mid \psi \rangle \simeq \langle \psi_{HF} \mid \hat{X} \mid \psi_{HF} \rangle ,
\end{equation}
for any one-body operator $\hat{X}$. 

The fully antisymmetrized wave functions are in 1-1 correspondence with the idempotent ($\rho^2=\rho$) density matrices $\rho_{\alpha^{\prime} \alpha} = \langle \psi_{HF} \mid a^{\dag}_{\alpha} a_{\alpha^{\prime}} \mid \psi_{HF} \rangle.$ The idempotent densities form a surface in the space of all density matrices that is invariant with respect to the coadjoint group action, $\rho \mapsto g\rho g^{-1}$ for $g\in U(n)$. The dimension of this coadjoint orbit surface is $2A(n-A)$ for the case of determinants -- a more manageable number than the factorially-growing dimension of the shell model space. The Hartree-Fock Hamiltonian 
\begin{equation}
H_{HF} = \sum_{\alpha \alpha^{\prime}} \left( T_{\alpha \alpha^{\prime}}  + \sum_{ \beta \beta^{\prime}} V_{\alpha \beta \alpha^{\prime} \beta^{\prime}} \rho_{\beta^{\prime} \beta} \right)  a^{\dag}_{\alpha} a_{\alpha^{\prime}}  
\end{equation}
is an element of the algebra $u(n)$ and it depends on the density.

The expectation of any one-body operator is given in terms of the density matrix by $\langle \hat{X} \rangle = tr(\rho X)$. Indeed the density matrix extracts from the microscopic wave function only the information about its single-particle degrees of freedom.  Thus, the Hartree-Fock approximation is adapted to provide the optimal description for the single-particle degrees of freedom of a many-fermion system. This is achieved mathematically by using the Lie algebra $u(n)$ of one-body operators. In a similar way Hartree-Fock-Bogoliubov theory has been presented as a coadjoint orbit of the group $O(2n)$ of quasiparticle transformations.~\cite{R81}

When other degrees of freedom dominate in an interacting many-body system, the method of coadjoint orbits should be generalized to the Lie algebra {\textbf g} corresponding to the physically relevant degrees of freedom. In this article the algebraic theory of mean fields is formulated and applied to the Elliott $SU(3)$ algebra. A synopsis of the generalization is as follows: The moment map establishes the connection between the Hilbert space of microscopic wave functions and the state densities. The densities are elements of the dual space {\textbf g}$^\ast$ to the Lie algebra. A generalized Hohenberg-Kohn theorem shows that there is an energy functional on the space of densities that is inherited from the microscopic Hamiltonian. Stationary densities can be obtained by minimizing the energy functional.  Each coadjoint orbit is an even dimensional phase space that has a nondegenerate symplectic structure. The symplectic form enables the mean field Hamiltonian, an element of the Lie algebra {\textbf g}, to be derived from the energy functional. Time-dependent mean field dynamics on the space of densities is a classical  dynamical system of Lax type.

\section{Moment Map and Coadjoint Orbits}
Denote the Hilbert space of $A$-fermion wave functions by $\textbf {H}$. Suppose a matrix Lie algebra {\textbf g} is represented faithfully by an algebra of hermitian fermion operators; the matrix $X\in$ {\textbf g} is represented by the hermitian operator $\hat{X}$ acting on $\textbf {H}$. Suppose G is the connected Lie group corresponding to the algebra {\textbf g} and ${\textbf U}$ is the corresponding faithful unitary representation of G on Fock space. The dual space {\textbf g}$^\ast$  is, by definition, the vector space of all real-valued linear functionals defined on {\textbf g}. The relation between the algebra and its dual space is similar to the one between contravariant and covariant vector spaces or between ket and bra  spaces. In the context of mean field theory the dual elements are the state densities and the notation $\rho$ will be used for them throughout this exposition. The bracket notation, $\langle \rho , X \rangle$, is used for the operation of the linear functional $\rho\in$ {\textbf g}$^\ast$ on the Lie algebra element $X\in$ {\textbf g}.

The moment map is defined as a mapping $M : \textbf {H} \rightarrow \mbox{\textbf g}^\ast$ from the Hilbert space to the dual of the Lie algebra that provides a unique density $\rho =  M({\psi})$ for every wave function $\psi\in{\textbf H}$,
\begin{equation}
\langle \rho , X \rangle = \frac {\langle \psi \mid \hat {X} \mid \psi \rangle}{\langle \psi \mid \psi \rangle}
\end{equation}
for all $X$ in {\textbf g}. Although the correspondence between wave functions and densities is one-to-one for the special case of Hartree-Fock, the moment map is many-to-one in the general case.

The moment  mapping is the heart of mean field theory. The density retains only part of the entire information about the system that the wave function carries, but a very important part -- the expectations of the observables that close the Lie algebra. Hence the moment map provides a significant simplification of the problem of how to describe a many-fermion system. It reduces all the degrees of freedom incorporated in the wave function to just those relevant to the observables of interest. If the algebra {\textbf g} contained all (not just one-body) operators, there would be no such reduction, the mapping would be one-to-one, and the shell model would be physically and mathematically equivalent to the mean field theory.

Another important feature of the moment map is that it induces an action of the Lie group on the dual space of densities from the unitary group representation in the Hilbert space. Given any wave function $\psi$ and its corresponding density $M(\psi)$, the group $G$ transforms the wave function into ${\textbf U}(g)\psi$ while the density is transformed into $M({\textbf U}(g) \psi)$. But the transformed density is given by
\begin{eqnarray}
\langle M({\textbf U}(g) \psi) \, , \, X \rangle \nonumber & = &  \frac {\langle {\textbf U}(g) \psi \mid \hat {X} \mid {\textbf U}(g) \psi \rangle}{\langle {\textbf U}(g) \psi \mid  {\textbf U}(g) \psi \rangle} \nonumber \\ [0.2cm]
& = &  \frac {\langle \psi \mid {\textbf U}(g)^{-1} \hat {X}  {\textbf U}(g) \mid \psi \rangle}{\langle \psi \mid \psi \rangle}  \nonumber \\ [0.2cm]
& = & \langle M(\psi) \, , \, g^{-1} X g \rangle ,
\end{eqnarray}
since the matrix $Ad_{g^{-1}}(X)\equiv g^{-1} X g$ is represented by the hermitian operator ${\textbf U}(g)^{-1} \hat {X}  {\textbf U}(g)$. By definition, the coadjoint group action $Ad^\ast$ on the space of densities is given by $\langle {{Ad_g}^\ast} \rho \, , \, X \rangle= \langle \rho \, , \, Ad_{g^{-1}} X \rangle$. Therefore it has been proven that ${{Ad_g}^\ast}(M(\psi)) = M({\textbf U}(g) \psi)$.

The mean field approximation restricts the allowed densities in the theory to a single coadjoint orbit. Fix a density $\rho\in$ {\textbf g}$^\ast$. The set of all densities that can be obtained by acting with the group in all possible ways on a given fixed density $\rho$ is its coadjoint orbit, 
\begin{equation}
{\textbf O}_{\rho} = \left\{ {Ad_g}^\ast \, \rho \in \mbox{\textbf g}^\ast \, \mid \, g \in G  \right\} .
\end{equation}
How does one work with coadjoint orbits? They are multidimensional surfaces and defining coordinates on them even locally is an extremely difficult task. Fortunately, there is a coordinate-free way. Coadjoint orbits are diffeomorphic to homogeneous spaces, i.e. they can be presented as coset spaces of the group modulo its isotropy subgroup. By definition the isotropy subgroup is the set of all group elements that fix a point $\rho\in$ {\textbf g}$^\ast$, 
\begin{equation}
G_\rho = \left\{ g \in G \mid {Ad_g^\ast} \rho = \rho \right\}.
\end{equation}
Then the coadjoint orbit is diffeomorphic to the homogeneous space 
\begin{equation}
{\textbf O}_{\rho} = G \, / \, G_\rho \, ,
\end{equation}
and its dimension is less than the dimension of the dual space itself, $\dim {\textbf O}_{\rho} = \dim G - \dim G_\rho = \dim$ {\textbf g}$^\ast - \dim G_\rho$.  This dimensional reduction is one more advantage to working with the coadjoint orbit mean field method. In Hartree-Fock the isotropy subgroup is the direct product $U(A)\times U(n-A)$ and the dimension of the coadjoint orbit of idempotent densities is $n^2-(A^2+(n-A)^2) = 2 A (n-A)$.

\section{Mean Field Dynamics}
The ground state density is found by minimizing the energy functional defined on the coadjoint orbit ${\textbf O}_{\rho}$. This energy functional should be derived from the microscopic Hamiltonian $\hat{H}$ on ${\textbf H}$. Because the moment map is many-to-one, the energy functional on the dual space cannot be set equal to the expectation of the Hamiltonian but instead is given by
\begin{equation}
{\textbf E}[\rho] =  \inf_{\psi \in M^{-1}(\rho)}  \frac {\langle \psi \mid \hat {H} \mid \psi \rangle}{\langle \psi \mid \psi \rangle} .
\end{equation}

The mean field Hamiltonian, an element of the Lie algebra {\textbf g}, is determined by both  the energy functional and the symplectic structure on the coadjoint orbit. The symplectic geometry is inherited from the Lie algebra as follows: Every Lie algebra element $X$ defines a tangent vector to the coadjoint orbit surface through the density $\rho$. This vector is the tangent to the curve $\gamma_{X}(t) = Ad^{\ast}_{g}(\rho) \in {\textbf O}_{\rho}$ where $g = \exp (t\,X)$. Note that the elements of the isotropy subalgebra {\textbf g}$_{\rho}$ correspond to null tangent vectors, and the tangent space is isomorphic to the coset vector space {\textbf g}/{\textbf g}$_{\rho}$. For any two tangent vectors $X, Y \in$ {\textbf g}/{\textbf g}$_{\rho}$ to the coadjoint orbit through $\rho$, the bilinear form,
\begin{equation}
\omega_{\rho}(X,Y) = \langle \rho, [X,Y] \rangle , 
\end{equation}
is antisymmetric, closed, and nondegenerate. Thus each coadjoint orbit is a symplectic manifold.~\cite{Kirillov} In particular, each surface ${\textbf O}_{\rho}$ is even dimensional. Moreover the group $G$ acts on each orbit as a transitive group of canonical transformations,
\begin{equation}
\omega_{Ad^{\ast}_{g}(\rho)}(Ad_{g}(X), Ad_{g}(Y)) = \omega_{\rho}(X,Y), 
\end{equation}  
where $X, Y$ are tangent vectors at $\rho$ and $Ad_{g}(X)$ and $Ad_{g}(Y)$ are tangent vectors at $Ad^{\ast}_{g}(\rho)$.

The model group theoretic Hamiltonian $h$ may be defined now. Choose a  coadjoint orbit ${\textbf O}_{\rho}$ in the dual space and let ${\textbf E}$ denote the energy functional for this orbit.  If $X$ is a tangent vector at $\rho$, denote the derivative of the energy in the $X$ direction by
\begin{equation}
d{\textbf E}_{\rho} (X) =  \frac{d}{dt} {\textbf E}(Ad_{g}^{\ast} (\rho) ) |_{t=0} ,\ \mbox{where\ } g = \exp(t\,X) .
\end{equation}
The model Hamiltonian $h\in$ {\textbf g}/{\textbf g}$_{\rho}$ is determined by
\begin{equation}
d{\textbf E}_{\rho} (X) = \omega_{\rho}(h,X) \mbox{\ for all } X \in \mbox{\textbf g} . \label{omega}
\end{equation}
There is a unique solution to this equation for $h\in$ {\textbf g}/{\textbf g}$_{\rho}$ because the symplectic form $\omega$ is nondegenerate on the tangent space. The model Hamiltonian $h$ depends on the point $\rho$ in the coadjoint orbit, as expected. The time evolution of a density on the coadjoint orbit surface can be shown to obey the Lax equation~\cite{Lax}
\begin{equation}
\dot {\rho} = [h(\rho) \, , \, \rho] .
\end{equation}
It can be easily shown that all the constants of motion for a Lax system are given by
\begin{equation}
I_n = \frac{1}{n} tr(\rho^n) .
\end{equation}

In the special case when the Hamiltonian $\hat{H}$ is an element of the Lie algebra {\textbf g}, the energy functional is just ${\textbf E}[\rho]=\langle \rho,H \rangle$ and the mean field Hamiltonian $h$ is equal to $H$. For Hartree-Fock, the model Hamiltonian $h$ on the coadjoint orbit of idempotent densities is identical to the usual mean field Hamiltonian. Otherwise the model Hamiltonian $h$ is the optimal approximation to the microscopic $\hat{H}$ relative to the Lie algebra {\textbf g} in the sense that $\langle \psi \, | \, [\hat{H}-h, X]\, \psi \rangle = 0$ for all $X \in$ {\textbf g} where $M(\psi)=\rho$.

\section{Mean Field Theory of the $SU(3)$ Model}
The angular momentum $L$ and the Elliott quadrupole operator ${\textbf Q}$ close under commutation to form the Lie  algebra $su(3)$.~\cite{JPE} Its irreducible representations are labeled by the pair of nonnegative integers $(\lambda \, , \, \mu)$, where $\lambda = n_{x} - n_{y}$, $\mu = n_{y} - n_{z}$, and $n_{x}$, $n_{y}$, $n_{z}$ are the oscillator quanta in the $x$, $y$, and $z$ directions of the highest weight vector $\psi_{HW}$. The Hamiltonian of the model is  
\begin{equation}
H_{su(3)} = H_{0} + (1/2) \, \kappa \, \textbf {Q} . \textbf {Q}
\end{equation}
with $H_{0}$ being the oscillator Hamiltonian and $\kappa < 0$. The energies in the spectrum are proportional to the eigenvalues of $L^2$.

The Elliott $su(3)$ algebra is a faithful representation of the complex antihermitian traceless $3\times 3$ matrices, 
\begin{equation}
su(3) = \left \{ \xi \in {\bf {M}}_{3}({\bf C}) \, \mid \, \xi ^\dag = - \xi \, , \, {\mbox tr} \, \xi = 0 \right \}.
\end{equation}
The real antisymmetric matrices correspond to the angular momentum and the pure imaginary symmetric matrices correspond to the quadrupole operator. Because $su(3)$ is a semisimple algebra, its Killing form is nondegenerate ( it can be shown that $\langle \rho , \xi \rangle = {\mbox tr}\,(\rho \, \xi)$ ), and therefore its dual space may be identified with the algebra itself
\begin{equation}
{su(3)}^\ast = \left \{ \rho \in {\bf {M}}_{3}({\bf C}) \, \mid \, \rho ^\dag = - \rho \, , \, {\mbox tr} \, \rho = 0 \right \} .
\end{equation}
If the density is expressed as a sum of real and imaginary parts, $\rho = A+iB$, then $\langle \rho , L_{ij} \rangle = A_{ij}$ and $\langle \rho, {\textbf Q}_{ij}\rangle = B_{ij}$.

The coadjoint action of the group $SU(3)$ on the dual space ${su(3)}^\ast$ is $Ad_g^\ast \, \rho = g \rho g^{-1}$, and the coadjoint orbit at each point $\rho \in su(3)^\ast$ is
\begin{equation}
{\textbf O}_{\rho} = \left \{ Ad_g^\ast \, \rho = g \rho g^{-1}  \, \mid \, g \in SU(3)  \right \} .
\end{equation}
Each orbit in this case contains a diagonal matrix with pure imaginary entries and can be written in terms of $(\lambda, \, \mu)$ as follows
\begin{equation} 
\rho_{HW} = \frac {i}{3} \left ( \begin{array}{ccc}
				2 \lambda + \mu  &              0          	&  0 \\
					0	        &  - \lambda + \mu       &  0 \\
					0	        &		0          & - \lambda - 2 \mu \\
\end{array} \right ) 
\end{equation}
Note that this diagonal density corresponds, via the moment map, to the highest weight vector $\psi_{HW}$. The coadjoint orbit is diffeomorphic to the coset space $G / G_{\rho}$, and the isotropy subgroup depends on the degeneracies of ${\rho}_{HW}$:
\begin{equation}
{\textbf O}_\rho = \left \{ \begin{array}{lcl}
	SU(3) / SU(3) = \left \{ \rho \right \} \, , & (\lambda \, , \, \mu) = (0 \, ,  0) \, , & \dim = 0 \\
	SU(3) / U(2) \, , & (\lambda \neq 0 \, , \, \mu = 0) {\mbox \ or\ } (\lambda=0 \, , \, \mu \neq 0) \, , & \dim = 4 \\
	SU(3) / U(1) \times U(1) \, , & (\lambda \neq 0 \, , \, \mu \neq 0) \, , & \dim = 6
\end{array} \right.
\end{equation}

Consider the quadrupole-quadrupole interaction $\textbf {Q} . \textbf {Q}$. The energy functional that corresponds to this interaction is ${\textbf E}[\rho] = {\mbox tr}\, B^2$. Using the symplectic structure the mean field approximation to the quadrupole-quadrupole interaction is,  according to Eq.\,(\ref{omega}), $h[\rho] = 2 \rho^t$, and the time evolution of the mean field is governed by the Lax equation
\begin{equation}
\dot{\rho} = [ 2\rho^t, \rho ] .
\end{equation}

For a Lax equation, it is easily shown that the trace of any power of $\rho$ is a constant of the motion. For the $su(3)$ highest weight orbit, the conserved trace of the square of the density is evaluated at the highest weight density to be $I_{2} \equiv  1/2\, {\mbox tr}\, \rho^2 = (\lambda^2+\lambda \mu+\mu^2)/3$. At an arbitrary orbit point $\rho=A+iB$, the quadratic Lax invariant is given by $I_{2} =  1/4 L^2 + 1/12\,{\textbf Q}\cdot{\textbf Q}$.


\end{document}